# Improving Schema Matching with Linked Data

Ahmad Assaf[†], Eldad Louw[†], Aline Senart[†], Corentin Follenfant[†],
Raphaël Troncy[‡] and David Trastour[†]

[†]SAP Research, SAP Labs France SAS
805 avenue du Dr. Maurice Donat, BP 1216
06254 Mougins Cedex, France
firstname.lastname@sap.com

[‡]EURECOM
BP193, F-06904 Sophia Antipolis Cedex, France
raphael.troncy@eurecom.fr

## ABSTRACT
With today's public data sets containing billions of data items, more and more companies are looking to integrate external data with their traditional enterprise data to improve business intelligence analysis. These distributed data sources however exhibit heterogeneous data formats and terminologies and may contain noisy data. In this paper, we present a novel framework that enables business users to semi-automatically perform data integration on potentially noisy tabular data. This framework offers an extension to Google Refine with novel schema matching algorithms leveraging Freebase rich types. First experiments show that using Linked Data to map cell values with instances and column headers with types improves significantly the quality of the matching results and therefore should lead to more informed decisions.

## 1. INTRODUCTION
Companies have traditionally performed business analysis based on transactional data stored in legacy relational databases. The enterprise data available for decision makers was typically relationship management or enterprise resource planning data [2]. However social media feeds, weblogs, sensor data, or data published by governments or international organizations are nowadays becoming increasingly available [3].

The quality and amount of structured knowledge available make it now feasible for companies to mine this huge amount of public data and integrate it in their next-generation enterprise information management systems. Analyzing this new type of data within the context of existing enterprise data should bring them new or more accurate business insights and allow better recognition of sales and market opportunities [4].

These new distributed sources however raise tremendous challenges. They have inherently different file formats, access protocols or query languages. They possess their own data model with different ways of representing and storing the data. Data across these sources may be noisy (e.g. duplicate or inconsistent), uncertain or be semantically similar yet different [5]. Integration and provision of a unified view for these heterogeneous and complex data structures therefore require powerful tools to map and organize the data.

In this paper, we present a framework that enables business users to semi-automatically combine potentially noisy data residing in heterogeneous silos. Semantically related data is identified and appropriate mappings are suggested to users. On user acceptance, data is aggregated and can be visualized directly or exported to Business Intelligence reporting tools. The framework is composed of a set of extensions to Google Refine server and a plug-in to its user interface [6]. Google Refine was selected for its extensibility as well as good cleansing and transformation capabilities [7].

We first map cell values with instances and column headers with types from popular data sets from the Linked Open Data Cloud. To perform the matching, we use the Auto Mapping Core (also called AMC [8]) that combines the results of various similarity algorithms. The novelty of our approach resides in our exploitation of Linked Data to improve the schema matching process. We developed specific algorithms on rich types from vector algebra and statistics. The AMC generates a list of high-quality mappings from these algorithms allowing better data integration.

First experiments show that Linked Data increases significantly the number of mappings suggested to the user. Schemas can also be discovered if column headers are not defined and can be improved when they are not named or typed correctly. Finally, data reconciliation can be performed regardless of data source languages or ambiguity. All these enhancements allow business users to get more valuable and higher-quality data and consequently to take more informed decisions.

The rest of the paper is organized as follows. Section 2 presents some related work. Section 3 describes the framework that we have designed for business users to combine data from heterogeneous sources. Section 4 validates our approach and shows the value of the framework through experiments. Finally, Section 5 concludes the paper and discusses future work.

## 2. RELATED WORK
While schema matching has always been an active research area in data integration, new challenges are faced today by the increasing size, number and complexity of data sources and their distribution over the network. Data sets are not always correctly typed or labeled and that hinders the matching process.

In the past, some work has tried to improve existing data schemas [9] but literature mainly covers automatic or semi-automatic labeling of anonymous data sets through Web extraction. Examples include [10] that automatically labels news articles with a tree structure analysis or [11] that defines heuristics based on distance and alignment of a data value and its label. These approaches are however restricting label candidates to Web content from which the data was extracted. [12] goes a step further by launching speculative queries to standard Web search engines to enlarge the set of potential candidate labels. More recently, [1] applies machine learning techniques to respectively annotate table rows as entities, columns as their types and pairs of columns as relationships, referring to the YAGO ontology. The work presented aims however at leveraging such annotations to assist semantic search queries construction and not at improving schema matching.

With the emergence of the Semantic Web, new work in the area has tried to exploit Linked Data repositories. The authors of [13]

present techniques to automatically infer a semantic model on tabular data by getting top candidates from Wikitology [14] and classifying them with the Google page ranking algorithm. Since the authors' goal is to export the resulting table data as Linked Data and not to improve schema matching, some columns can be labeled incorrectly, and acronyms and languages are not well handled [13]. In the Helix project [15], a tagging mechanism is used to add semantic information on tabular data. A sample of instances values for each column is taken and a set of tags with scores are gathered from online sources such as Freebase [16]. Tags are then correlated to infer annotations for the column. The mechanism is quite similar to ours but the resulting tags for the column are independent of the existing column name and sampling might not always provide a representative population of the instance values.

## 3. PROPOSITION

Google Refine (formerly Freebase Gridworks) is a tool designed to quickly and efficiently process, clean and eventually enrich large amounts of data with existing knowledge bases such as Freebase. The tool has however some limitations: it was initially designed for data cleansing on only one data set at a time, with no possibility to compose columns from different data sets. In this section, we describe in detail our framework allowing data mashup from several sources. We first present our framework architecture, then the activity flow and finally our approach to schema matching.

### 3.1 Framework Architecture

Google Refine makes use of a modular web application framework similar to OSGi called Butterfly [17]. The server-side written in Java maintains states of the data (undo/redo history, long-running processes, etc.) while the client-side implemented in Javascript maintains states of the user interface (facets and their selections, view pagination, etc.). Communication between the client and server is done through REST web services.

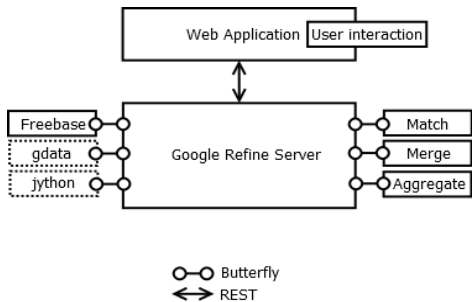

**Figure 1. Framework Architecture**

As depicted in Figure 1, our framework leverages Google Refine and defines three new Butterfly modules to extend the server's functionality (namely Match, Merge and Aggregate modules) and one JavaScript extension to capture user interaction with these new data matching capabilities.

### 3.2 Activity Flow

This section presents the sequence of activities and interdependencies between these activities when using our framework. Figure 2 gives an outline of these activities.

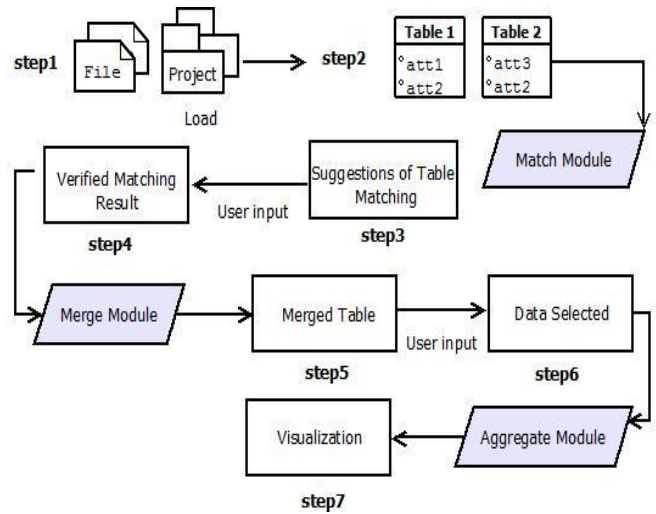

**Figure 2. Activity Flow**

The data sets to match can be contained in files (e.g. csv, Excel spreadsheets, etc.) or defined in Google Refine projects (step 1). The inputs for the match module are the source and target files and/or projects that contain the data sets. These projects are imported into the internal data structure (called schema) of the AMC [18] (step 2). The AMC then uses a set of built-in algorithms to calculate similarities between the source and target schemas on an element basis, i.e. column names in the case of spreadsheets or relational databases. The output is a set of similarities, each containing a triple consisting of source schema element, target element, and similarity between the two. As depicted in Figure 3, these results are presented to the user in tabular form (step 3) such that s/he can check, correct, and potentially complete them (step 4).

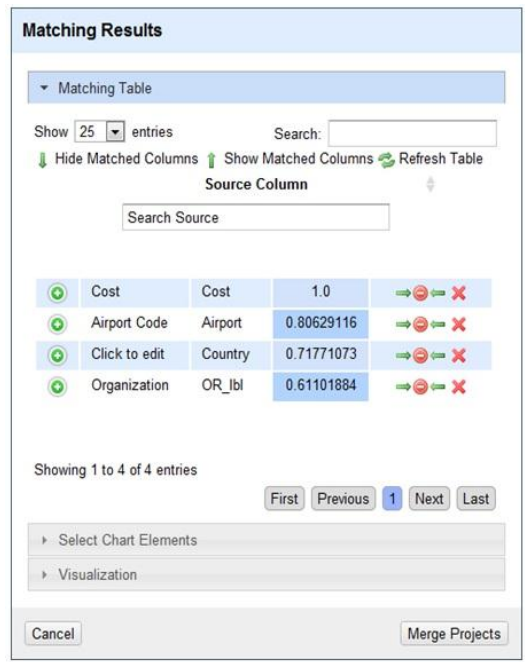

**Figure 3. Suggestions of Table Matching**

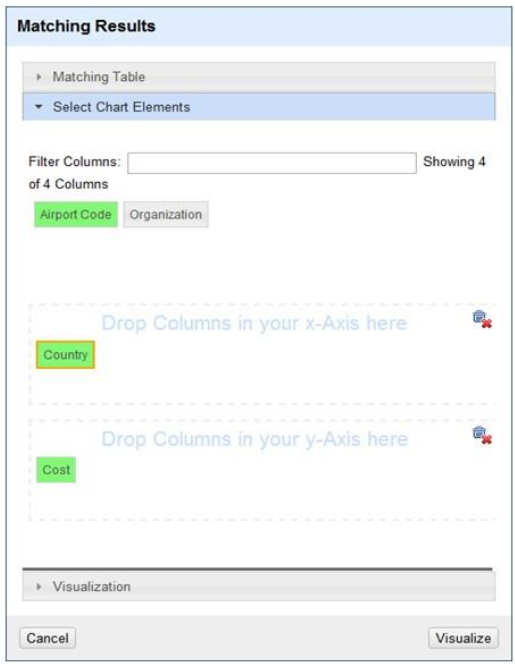

**Figure 4. Data Selection**

Once the user has completed the matching of columns, the merge information is sent back to Google Refine, which calls the merge module. This module creates a new project, which contains a union of the two projects where the matched columns of the target project are appended to the corresponding source columns (step 5). As shown in Figure 4, the user can then select the columns that s/he wants to merge and visualize by dragging and dropping the required columns onto the fields that represent the x and y axes (step 6).

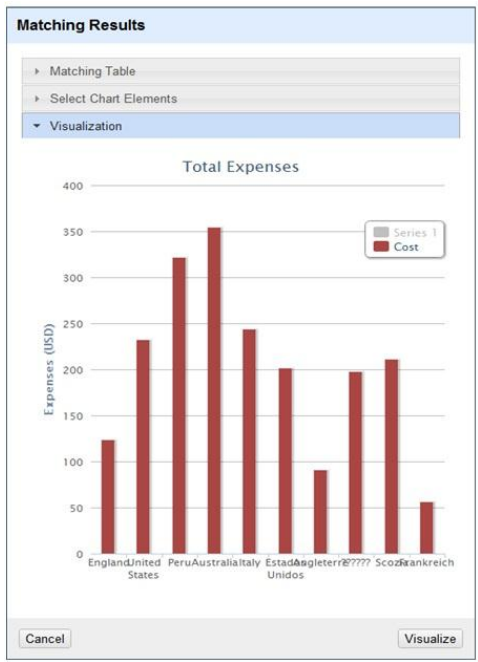

**Figure 5. Data Visualization**

Once the selection has been performed, the aggregation module merges the filtered columns and the result can then be visualized as shown in Figure 5 (step 7). As aggregation operations can quickly become complex, our default aggregation module can be replaced by more advanced analytics on tabular data. The integration of such a tool is part of future work.

### 3.3 Schema Matching

Schema matching is typically used in business to business integration, metamodel matching, as well as Extract, Transform, Load (ETL) processes. For non-IT specialists the typical way of comparing financial data from two different years or quarters, for example, would be to copy and paste the data from one Excel spreadsheet into another one, thus creating redundancies and potentially introducing copy-and-paste errors. By using schema matching techniques it is possible to support this process semi-automatically, i.e. to determine which columns are similar and propose them to the user for integration. This integration can then be done with appropriate business intelligence tools to provide visualisations.

One of the problems in performing the integration is the quality of data. The columns may contain data that is noisy or incorrect. There may also be no column headers to provide suitable information for matching. A number of approaches exploit the similarities of headers or similarities of types of column data. We propose a new approach that exploits semantic rich typing provided by popular datasets from the Linked Data cloud.

#### 3.3.1 Data Reconciliation

Reconciliation enables entity resolution, i.e. matching cells with corresponding typed entities in case of tabular data. Google Refine already supports reconciliation with Freebase but requires confirmation from the user. For medium to large data sets, this can be very time-consuming. To reconcile data, we therefore first identify the columns that are candidates for reconciliation by skipping the columns containing numerical values or dates. We then use the Freebase search API to query for each cell of the source and target columns the list of typed entities candidates. Results are cached in order to be retrieved by our similarity algorithms.

#### 3.3.2 Matching Unnamed and Untyped Columns

The AMC has the ability to combine the results of different matching algorithms. Its default built-in matching algorithms work on column headers and produce an overall similarity score between the compared schema elements. It has been proven that combining different algorithms greatly increases the quality of matching results [8] [19]. However, when headers are missing or ambiguous, the AMC can only exploit domain intersection and inclusion algorithms based on column data. We have therefore implemented three new similarity algorithms that leverage the rich types retrieved from Linked Data in order to enhance the matching results of unnamed or untyped columns. They are presented below.

##### 3.3.2.1 Cosine Similarity

The first algorithm that we implemented is based on vector algebra. Let $v$ be the vector of ranked candidate types returned by Freebase for each cell value of a column. Then:

$$v := \sum_{i=1}^{K} a_i * \vec{t_i}$$

where $a_i$ is the score of the entry and $\vec{t_i}$ is the type returned by Freebase. The vector notation is chosen to indicate that each distinct answer determines one dimension in the space of results.

Each cell value now has a weighted result set that can be used for aggregation to produce a result vector for the whole column. The column result $V$ is then given by:

$$V := \sum_{i=1}^{n} v_i$$

We now compare the result vector of candidate types from the source column with the result vector of candidate types from the target column. Let $W$ be the result vector for the target column, then the similarity $s$ between the columns pair can be calculated using the absolute value of the cosine similarity function:

$$s := \frac{|(V * W)|}{\|V\| * \|W\|}$$

### 3.3.2.2 Pearson Product-Moment Correlation Coefficient (PPMCC)

The second algorithm that we implemented is PPMCC, a statistical measure of the linear independence between two variables $(x, y)$ [20]. In our method, x is an array that represents the total scores for the source column rich types, y is an array that represents the mapped values between the source and the target columns. The values present in x but not in y are represented by zeros. We have:

$SourceColumn\ [\{R_1, C_{sr1}\}, \{R_2, C_{sr2}\}, \{R_3, C_{sr3}\} \ldots \{R_n, C_{srn}\}]$

$TargetColumn\ [\{R_1, C_{tr1}\}, \{R_2, C_{tr2}\}, \{R_3, C_{tr3}\} \ldots \{R_n, C_{trn}\}]$

Where $R_1, R_2, \ldots, R_n$ are different rich type values retrieved from Freebase, $C_{sr1}, C_{sr2}, \ldots, C_{srn}$ are the sum of scores for each corresponding r occurrence in the source column, and $C_{tr1}, C_{tr2}, \ldots, C_{trn}$ are the sum of scores for each corresponding r occurrence in the target column.

The input for PPMC consists of two arrays that represent the values from the source and target columns, where the source column is the column with the largest set of rich types found. For example:

$$X = [C_{sr1}, C_{sr2}, C_{sr4}, \ldots, C_{srn}]$$

$$Y = [0, C_{tr2}, C_{tr4}, \ldots, C_{trn}]$$

Then the sample correlation coefficient (r) is calculated using:

$$r = \frac{\sum_{i=1}^{n}(x_i - \bar{x})(y_i - \bar{y})}{\sqrt{\sum_{i=1}^{n}(x_i - \bar{x})^2} \sqrt{\sum_{i=1}^{n}(y_i - \bar{y})^2}}$$

Based on a sample paired data $(x_i, y_i)$, the sample PPMCC is:

$$r = \frac{1}{n-1}\sum_{i=1}^{n}\left(\frac{x_i - \bar{x}}{s_x}\right)\left(\frac{y_i - \bar{y}}{s_y}\right)$$

Where $\left(\frac{x_i - \bar{x}}{s_x}\right)$, $\bar{x}$ and $s_x$ are the standard score, sample mean and sample standard deviation, respectively.

### 3.3.2.3 Spearman's Rank Correlation Coefficient

The last algorithm that we implemented to match unnamed and untyped columns is Spearman's rank correlation coefficient [21]. It applies a rank transformation on the input data and computes PPMCC afterwards on the ranked data. In our experiments we used Natural Ranking with default strategies for handling ties and NaN values. The ranking algorithm is however configurable and can be enhanced by using more sophisticated measures.

### 3.3.3 Column Labeling

We showed in the previous section how to match unnamed and untyped columns. Column labeling is however beneficial as the results of our previous algorithms can be combined with traditional header matching techniques to improve the quality of matching.

Rich types retrieved from Freebase are independent from each other. We need to find a method that will determine normalized score for each type in the set by balancing the proportion of high scores with the lower ones. We used Wilson score interval for a Bernoulli parameter that is presented in the following equation:

$$w = \left(\hat{p} + \frac{z_{\alpha/2}^2}{2n} \pm z_{\alpha/2}\sqrt{\left[\hat{p}(1-\hat{p}) + \frac{z_{\alpha/2}^2}{4n}\right]/n}\right) / (1 + z_{\alpha/2}^2/n)$$

Here $\hat{p}$ is the average score for each rich type, n is the total number of scores and $z_{\alpha/2}$ is the score level; in our case it is 1.96 to reflect a score level of 0.95.

## 4. FIRST EXPERIMENTS

We present in this section early results from experiments we conducted using the different methods described above. To appreciate the value of our approach, please consider the two simple Excel spreadsheets in Table 1 and Table 2:

| Airport Code | | Organization | Cost |
|---|---|---|---|
| LHR | England | Microsoft | 123.2 |
| LGA | United States | Apple | 232.12 |
| HUU | Peru | Orange | 321.7 |
| DBO | Australia | IBM | 354.64 |
| BGY | Italy | Accenture | 243.8 |

**Table 1. Source Table**

| Airport | Pays | OR_lbl | Cost |
|---|---|---|---|
| LaGuardia | Estados Unidos | MS | 201.41 |
| Heathrow | Angleterre | Yahoo | 90.5 |
| Queen Alia | الأردن | Samsung | 198 |
| Prestwick | Scozia | GOOG | 211.27 |
| Beauvais | Frankreich | HP | 55.99 |

**Table 2. Target Table**

Most of the column headers in the source table exist and adequately present the data. The language is English, and airports are represented by their IATA code. The target table presents another set of data which has been produced by combining multiple queries from different data sources. As you can see, the country column is labeled in French while the values are written in different languages (Italian, Spanish, German, French and Arabic). The organization column is code-labeled and companies are either represented by their full name or by their NASDAQ code.

Running AMC with its default matchers returns the matching results shown in Table 3.

| Source Column | Target Column | Similarity |
|---|---|---|
| Cost | Cost | 1 |
| Airport Code | Airport | 0.7142857 |

Table 3. Similarity Scores Using the AMC Default Matching Algorithms

The AMC has perfectly matched the two columns labeled "Cost" using name and data type similarity calculations. Moreover, it has computed a similarity of approximately 71% between the "Airport Code" and "Airport" columns. However, there is no alignment found between the other columns since their headers are not related to each other, although the actual values are similar.

The Cosine Similarity algorithm combined with the AMC default matchers produces the results shown in Table 4.

| Source Column | Target Column | Similarity |
|---|---|---|
| Cost | Cost | 1 |
| Airport Code | Airport | 0.7741357 |
| Click to edit | Country | 0.7024157 |

Table 4. Enhanced Similarity Scores Adding the Cosine Similarity Method

We notice that the similarity score for the "Airport" column has increased slightly, and that the "Country" column is aligned to the blank header. This shows that our approach allows performing schema matching on columns with no headers.

The similarity score is an average of the applied algorithms (AMC's native and Cosine). The relatively high similarity score of "Country" column is explained by the fact that the native AMC matching algorithm has skipped that column as it does not have a valid header, and the results are solely those of the Cosine matcher. Likewise, the Cosine matcher skips checking the "Cost" columns as they contain numeric values, so only the AMC's native matcher results are taken into account. Finally, the "Organization" column is still not mapped as similarities under a threshold of 50% are ignored.

The second matching algorithm (PPMCC) combined with the previous algorithm yields the results presented in Table 5.

| Source Column | Target Column | Similarity |
|---|---|---|
| Cost | Cost | 1 |
| Airport Code | Airport | 0.80629116 |
| Click to edit | Country | 0.7177106 |
| Organization | OR_lbl | 0.61101884 |

Table 5. Enhanced Similarity Scores Adding the PPMCC Method

We now notice enhanced similarity scores and higher number of mappings. Mainly, the "Oganization" column from the source table has being aligned to the correct corresponding column "OR_lbl".

The third matching algorithm (Spearman's matcher) combined with the two previous ones generates the final results in Table 6.

| Source Column | Target Column | Similarity |
|---|---|---|
| Cost | Cost | 1 |
| Click to edit | Country | 0.66648626 |
| Airport Code | Airport | 0.6370289 |
| Organization | OR_lbl | 0.5439194 |

Table 6. Similarity Scores Using the Combination of Algorithms

The similarity results have slightly decreased when plugging Spearman's matcher. Several experiments have shown that this method does not work well with noisy data sets. For instance, the similarity results returned by Cosine, Pearson's and Spearman's matchers for the {Airport, Airport Code} pair is much higher: 83%, 87% and 13% respectively.

In a second set of experiments, we compare our previous results with less noisy data sets. Consider the tables 7 and 8.

| pays | organization | country | |
|---|---|---|---|
| Uganda | ibm | france | accenture |
| Zimbabwe | microsoft | iran | microsoft |
| Iran | google | Iran | google |
| Iraq | sap | jordan | sap |
| Libya | orange | england | orange |
| Syria | apple | Syria | apple |

Table 7. Source Table     Table 8. Target Table

Running the AMC on these tables will fail as their column headers strictly do not match. Combining Cosine and Pearson's methods with the AMC's native matchers' results in:

| organization | Click to edit | 0.726031 |
|---|---|---|
| Pays | country | 0.600207 |

Table 9. Enhanced Similarity Scores Adding the Cosine and Pearson's Method

Adding Pearson's method was found to enhance the results when dealing with relatively clean data sets. For the above example the result obtained is:

| organization | Click to edit | 0.78427 |
|---|---|---|
| Pays | country | 0.643885 |

Table 10. Similarity Scores Using the Combination of Algorithms

For the {Pays, country} pair, the similarity results returned by Cosine, Pearson's and Spearman's matchers are 99.3%, 99%, 95.8% respectively. Therefore, using the AMC allows identifying the best matching algorithms for a given data set.

Finally, applying our labeling method on the above data sets suggested relevant column names. For instance, looking at the unlabeled column, the system suggested "Organization" with a score of 1.72 compared to the next top score which is "Organism Classification" with a score of 0.371.

## 5. CONCLUSION AND FUTURE WORK

In this paper, we presented a framework enabling mashup of potentially noisy enterprise and external data. The implementation is based on Google Refine and uses Freebase to annotate data with rich types. As a result, the matching process of heterogeneous data sources is improved. Our preliminary evaluation shows that for data sets where mappings were relevant yet not proposed, our framework provides higher quality matching results. Additionally, the number of matches discovered is increased when Linked Data is used in most data sets. We plan in future work to evaluate the framework on larger data sets using rigorous statistical analysis of [22]. We also consider integrating additional linked open data sources of semantic types such as DBpedia [23] or YAGO [24] and evaluate our matching results against instance-based ontology alignment benchmarks such as [25] or [26]. Another future work will be to generalize our approach on data schemas to data classification. The same way the AMC helps identifying the best matches for two datasets, we plan to use it for identifying the best statistical classifiers for a sole dataset, based on normalized scores.

## 6. REFERENCES

[1] Girija Limaye, Sunita Sarawagi, and Soumen Chakrabarti, "Annotating and Searching Web Tables Using Entities, Types and Relationships," *Proceedings of the VLDB Endowment*, vol. III, no. 1, pp. 1338-1347, September 2010.

[2] Michael James Hernandez, *Database design for mere mortals: a hands-on guide to relational database design*.: Addison-Wesley, 2003.

[3] Danah Boyd and Kate Crawford, "Six Provocations for Big Data," *Computer and Information Science*, vol. 123, no. 1, 2011.

[4] Steve LaValle, Eric Lesser, Rebecca Shockley, Michael S. Hopkins, and Nina Kruschwitz, "Big Data, Analytics and the Path from Insights to Value," *MIT Sloan Management Review*, vol. 52, no. 2, 2011.

[5] C. Kavitha, G. Sudha Sadasivam, and Sangeetha N. Shenoy, "Ontology Based Semantic Integration of Heterogeneous Databases," *European Journal of Scientific Research*, vol. 64, no. 1, pp. 115-122, 2011.

[6] Google Code. Google Refine. [Online]. http://code.google.com/p/google-refine/

[7] Christian Bizer, Tom Heath, and Tim Berners-Lee, "Linked Data - The Story So Far," *International Journal on Semantic Web and Information Systems*, vol. 5, no. 3, pp. 1-22, 2009.

[8] Eric Peukert, Julian Eberius, and Rahm Erhard, "A Self-Configuring Schema Matching System," in *28th IEEE International Conference on Data Engineering*, 2012.

[9] Renee J. Miller and Periklis Andritsos, "On Schema Discovery," *IEEE Data Engineering Bulletin*, vol. 26, no. 3, pp. 40-45, 2003.

[10] Davy de Castro Reis, Paulo B. Golgher, Altigran S. da Silva, and Alberto H. F. Laender, "Automatic Web News Extraction Using Tree Edit Distance," in *13th International Conference on World Wide Web*, 2004.

[11] Jiying Wang and Fred Lochovsky, "Data Extraction and Label Assignment for Web Databases," in *12th International Conference on World Wide Web*, 2003.

[12] Altigran S. da Silva, Denilson Barbosa, M. B. Joao Cavalcanti, and A. S. Marco Sevalho, "Labeling Data Extracted from the Web," in *International Conference on the Move to Meaningful Internet Systems*, 2007.

[13] Tim Finin, Zareen Syed, Varish Mulwad, and Anupam Joshi, "Exploiting a Web of Semantic Data for Interpreting Tables," in *Web Science Conference*, 2010.

[14] Tim Finin, Zareen Syed, James Mayfield, Paul McNamee, and Christine Piatko, "Using Wikitology for Cross-Document Entity Coreference Resolution," in *AAAI Spring Symposium on Learning by Reading and Learning to Read*, 2009.

[15] Oktie Hassanzadeh et al., "Helix: Online Enterprise Data Analytics," in *20th International World Wide Web Conference - Demo Track*, 2011.

[16] Metaweb Technologies. Freebase. [Online]. http://www.freebase.com/

[17] Google Code. Smilie Butterfly. [Online]. http://code.google.com/p/simile-butterfly/

[18] Eric Peukert, Julian Eberius, and Erhard Rahm, "AMC - A Framework for Modelling and Comparing Matching Systems as Matching Processes," in *International Conference on Data Engineering - Demo Track*, 2011.

[19] Umberto Straccia and Raphael Troncy, "oMAP: Combining Classifiers for Aligning Automatically OWL Ontologies," in *6th International Conference on Web Information Systems Engineering*, 2005, pp. 133-147.

[20] Charles J. Kowalski, "On the Effects of Non-Normality on the Distribution of the Sample Product-Moment Correlation Coefficient," *Journal of the Royal Statistical Society*, vol. 21, no. 1, pp. 1-12, 1972.

[21] Sarah Boslaugh and Paul Andrew Watters, *Statistics in a Nutshell*.: O'Reilly Media, 2008.

[22] Tom Fawcett, "An Introduction to ROC Analysis," *Journal of Pattern Recognition Letters*, vol. 27, no. 8, 2006.

[23] Soren Auer et al., "DBpedia: A Nucleus for a Web of Open Data," in *6th International and 2nd Asian Semantic Web Conference* , 2007.

[24] Fabian M. Suchanek, Gjergji Kasneci, and Gerhard Weikum, "Yago: a Core of Semantic Knowledge," in *16th International Conference on World Wide Web*, 2007.

[25] (2011) Instance Matching at OAEI. [Online]. http://oaei.ontologymatching.org/2011/instance/index.html

[26] Alfio Ferrara. ISLab Instance Matching Benchmark. [Online]. http://islab.dico.unimi.it/iimb/